\documentclass[preprint,amsmath,amssymb,11pt,english,aps,showpacs,showkeys]{revtex4}
\usepackage{graphicx}
\usepackage{graphics}
\usepackage{amsmath}
\usepackage{dcolumn}
\usepackage{amssymb}
\usepackage{bm}
\usepackage[latin1]{inputenc}

\begin{document}
\title{Rotating effects on the Dirac oscillator in the cosmic string spacetime}
\author{K. Bakke}
\email{kbakke@fisica.ufpb.br}
\affiliation{Departamento de F\'isica, Universidade Federal da Para\'iba, Caixa Postal 5008, 58051-970, Jo\~ao Pessoa, PB, Brazil.}

\begin{abstract}
In this contribution, we study the Dirac oscillator under the influence of noninertial effects of a rotating frame in the cosmic string spacetime. We show that both noninertial effects and the topology of the cosmic string spacetime restrict the physical region of the spacetime where the quantum particle can be placed, and discuss two different cases of bound states solutions of the Dirac equation by analysing the behaviour of the Dirac oscillator frequency.
\end{abstract}


\maketitle

\section{Introduction}

In recent decades, the Dirac oscillator \cite{osc1,osc2} has attracted a great interest in studies of the Jaynes-Cummings model \cite{jay2,osc3}, the Ramsey-interferometry effect \cite{osc6} and quantum phase transitions \cite{extra2,extra3}. In Ref. \cite{jay2}, an exact mapping of the Dirac oscillator onto the Jaynes-Cummings model \cite{jay} has achieved a description of the interaction of a two-level atom with a quantized single-mode field. Besides, the results of Ref. \cite{osc3} have shown that the entanglement of the spin with the orbital motion is produced in a way similar to what is observed in the model of the Jaynes-Cummings model when the limit the strong spin-orbit coupling of the Dirac oscillator is considered. 

The Dirac oscillator was proposed in Ref. \cite{osc1} with the aim of recovering the harmonic oscillator Hamiltonian from the nonrelativistic limit of the Dirac equation. In short, by considering the nonrelativistic limit of the Dirac equation describing the interaction between a Dirac particle with the harmonic oscillator potential, the impossibility of recovering the harmonic oscillator Hamiltonian occurs due to the presence of a quadratic potential \cite{osc1,do2,osc2} in the second order differential equation. Therefore, in order to solve this problem, the relativistic harmonic oscillator has been investigated recently by introducing scalar and vector potentials which are quadratics in coordinates \cite{prc} and by introducing a new coupling into the Dirac equation \cite{osc1}. In particular, the new coupling proposed in Ref. \cite{osc1} is introduced in such a way that the Dirac equation remains linear in both spatial coordinates and momenta, and recover the Schr\"odinger equation for a harmonic oscillator in the nonrelativistic limit of the Dirac equation. The Dirac oscillator has also been analyzed in a series of physical systems, such as in a thermal bath \cite{osc10}, in the presence of an external magnetic field \cite{osc9}, in the point of view of the Lie algebra \cite{quesnemo}, in studies of covariance properties \cite{moreno}, in the hidden supersymmetry \cite{benitez,moreno,quesne2}, by using the shape invariant method \cite{quesne}, in the presence of the Aharonov-Bohm quantum flux \cite{ferk,luis,rojas,victor}, conformal invariance properties \cite{romerom}, in the Fermi-Walker reference frame \cite{b10} and for a system of a charged particle interacting with a topological defect \cite{josevi}. Recently, a model of a quantum ring based on the Dirac oscillator has been proposed to confine relativistic spin-$1/2$ particles \cite{bf30}.   

The aim of this paper is the study the Dirac oscillator under the influence of noninertial effects of a rotating frame in the cosmic string spacetime. Studies of noninertial effects have discovered interesting quantum effects such as the Sagnac effect \cite{sag,sag5}, the Mashhoon effect \cite{r3}, persistent currents in quantum rings \cite{r11}, the coupling between the angular momentum and the angular velocity of the rotating frame \cite{r1,r2,r4} and spin currents \cite{r12}. Other works that worth mentioning are studies involving rotational and gravitational effects in quantum interference \cite{r13,r14,r15}, Dirac fields \cite{r10}, scalar fields \cite{r8}, the influence of Lorentz transformations \cite{r5}, the weak field approximation \cite{r6}, the Landau-He-McKellar-Wilkens quantization \cite{b} and confinement of a neutral particle to a quantum dot \cite{b4,b12}. In this work, we show that both noninertial effects and the topology of the cosmic string spacetime restrict the physical region of the spacetime where the quantum particle can be placed. By analysing the behaviour of the oscillator frequency, we discuss two different cases of bound states solutions of the Dirac equation.

The structure of this paper is: in section II, we introduce the Dirac oscillator \cite{osc1,osc2} and discuss the influence of the noninertial effects of a rotating frame on the Dirac oscillator in the cosmic string spacetime. We show the restriction imposed by both noninertial effects and the topology of the cosmic string spacetime on the physical region of the spacetime where the quantum particle can be placed and analyse two different cases of bound states solutions of the Dirac equation; in section III, we present our conclusions.

\section{Dirac oscillator in a rotating frame in a topological defect spacetime}

We start this section by introducing the Dirac oscillator \cite{osc1,osc2}. In the following, we introduce the classical background of this work, build a rotating frame and discuss the influence of noninertial effects of a rotating frame and the topology of the classical background on the Dirac oscillator. Finally, we analyse the behaviour of the oscillator frequency by showing two different cases of bound states solutions of the Dirac equation.

As we have pointed out in the introduction, the term ``Dirac oscillator'' was denominated by Moshinsky and Szczepaniak \cite{osc1} based on the introduction of a coupling in such a way that the Dirac equation remains linear in both spatial coordinates and momenta. This new coupling introduced into the Dirac equation is given by 
\begin{eqnarray}
\vec{p}\rightarrow\vec{p}-im\omega\rho\,\hat{\beta}\,\hat{\rho},
\label{1.1}
\end{eqnarray}
where $m$ is the mass of the Dirac neutral particle, $\omega$ is the oscillator frequency, $\hat{\beta}$ is one of the standard Dirac matrices, $\rho=\sqrt{x^{2}+y^{2}}$ is the radial coordinate and $\hat{\rho}$ is a unit vector in the radial direction. 

Let us introduce the scenario of general relativity which corresponds to the cosmic string spacetime. The cosmic string spacetime is described by a line element given by \cite{vil,kibble} (we work with the units $\hbar=c=1$)
\begin{eqnarray}
ds^{2}=-dT^{2}+dR^{2}+\eta^{2}R^{2}d\Phi^{2}+dZ^{2},
\label{1.2}
\end{eqnarray}
where the parameter $\eta$ is related to the deficit angle which is defined as $\eta=1-4\varpi$, with $\varpi$ being the linear mass density of the cosmic string. In the cosmic string spacetime, the parameter $\eta$ can assume only values in which $\eta<1$. All values $\eta>1$ would correspond to a spacetime with negative curvature which does not make sense in the general relativity context \cite{kat,furt,moraesG2}. The azimuthal angle varies in the interval: $0\leq\varphi<2\pi$. Besides, the geometry described by the line element (\ref{1.2}) possesses a conical singularity represented by the curvature tensor
\begin{eqnarray}
\label{curv}
R_{\rho,\varphi}^{\rho,\varphi}=\frac{1-\eta}{4\eta}\,\delta_{2}(\vec{r}),
\label{1.3}
\end{eqnarray}
where $\delta_{2}(\vec{r})$ is the two-dimensional delta function, and we have considered the units $\hbar=c=1$. The behaviour of the curvature tensor (\ref{1.3}) is denominated as a conical singularity \cite{staro} which gives rise to the curvature concentrated on the cosmic string axis with a null curvature in all other points of the spacetime. In recent decades, a great deal of works have discussed the influence of the topology of the cosmic string spacetime, for instance, in the problem of string sources in $\left(2+1\right)$-dimensional gravity \cite{extra1}, in the Kaluza-Klein theory \cite{kaluza}, in the gravitational analogue of the Aharonov-Bohm effect \cite{hol}, in the Fermi-Walker transport \cite{anan3}, in the relativistic quantum scattering \cite{scat}, in the Lorentz symmetry violation background \cite{ed} and the Aharonov-Bohm effect for bound states \cite{val}.

Now, let us make a coordinate transformation: $T=t$, $R=\rho$, $\Phi=\varphi+\omega\,t$ and $Z=z$, where $\omega$ is the constant angular velocity of the rotating frame. Thereby, the line element (\ref{1.2}) becomes
\begin{eqnarray}
ds^{2}=-\left(1-\omega^{2}\eta^{2}\rho^{2}\right)\,dt^{2}+2\omega\eta^{2}\rho^{2}d\varphi\,dt+d\rho^{2}+\eta^{2}\rho^{2}d\varphi^{2}+dz^{2}.
\label{1.4}
\end{eqnarray}

We can note that the line element (\ref{1.4}) is defined for values of the radial coordinate inside the range: 
\begin{eqnarray}
0\,<\rho\,<\,1/\omega\eta.
\label{1.5}
\end{eqnarray}
Therefore, all values for the radial coordinate given by $\rho\,>\,1/\omega\eta$ mean that the particle is placed outside of the light-cone because the velocity of the particle is greater than the velocity of the light \cite{landau3}. In this way, the range (\ref{1.5}) imposes a spatial constraint where the wave function must be defined. This restriction on on the radial coordinate imposed by both noninertial effects and the topology of the cosmic string spacetime gives rise to a hard-wall confining potential, that is,  it imposes that the wave function of the Dirac particle must vanish at $\rho\rightarrow1/\omega\eta$.

Further, in order to work with Dirac spinors in a rotating frame in a topological defect spacetime having a non-null curvature concentrated on the symmetry axis, we use the formulation of spinors in curved spacetime \cite{weinberg,bd}. In a curved spacetime background, spinors are defined locally. Each spinor transforms according to the infinitesimal Lorentz transformations, that is, $\psi'\left(x\right)=D\left(\Lambda\left(x\right)\right)\,\psi\left(x\right)$, where $D\left(\Lambda\left(x\right)\right)$ corresponds to the spinor representation of the infinitesimal Lorentz group and $\Lambda\left(x\right)$ corresponds to the local Lorentz transformations \cite{weinberg}. Locally, the reference frame of the observers can be build via a noncoordinate basis $\hat{\theta}^{a}=e^{a}_{\,\,\,\mu}\left(x\right)\,dx^{\mu}$, where the components $e^{a}_{\,\,\,\mu}\left(x\right)$ are called \textit{tetrads} and satisfy the relation \cite{,weinberg,bd,naka}: $g_{\mu\nu}\left(x\right)=e^{a}_{\,\,\,\mu}\left(x\right)\,e^{b}_{\,\,\,\nu}\left(x\right)\,\eta_{ab}$, where $\eta_{ab}=\mathrm{diag}(- + + +)$ is the Minkowski tensor. The inverse of tetrads is defined as $dx^{\mu}=e^{\mu}_{\,\,\,a}\left(x\right)\,\hat{\theta}^{a}$, where we have the relations: $e^{a}_{\,\,\,\mu}\left(x\right)\,e^{\mu}_{\,\,\,b}\left(x\right)=\delta^{a}_{\,\,\,b}$ and $e^{\mu}_{\,\,\,a}\left(x\right)\,e^{a}_{\,\,\,\nu}\left(x\right)=\delta^{\mu}_{\,\,\,\nu}$. Observe that the indices $\mu,\nu\ldots$ indicate the spacetime indices, while the indices $a,b,c=0,1,2,3$ indicate the local reference frame. From this perspective, let us consider the following local reference frame for the observers \cite{bf4,bf16}:
\begin{eqnarray}
e^{a}_{\,\,\,\mu}\left(x\right)=\left(
\begin{array}{cccc}
\sqrt{1-\beta^{2}} & 0 & -\frac{\omega\eta^{2}\rho^{2}}{\sqrt{1-\beta^{2}}} & 0 \\
0 & 1 & 0 & 0 \\
0 & 0 & \frac{\eta\rho}{\sqrt{1-\beta^{2}}} & 0 \\
0 & 0 & 0 & 1 \\
\end{array}\right);\,\,\,
e^{\mu}_{\,\,\,a}\left(x\right)=\left(
\begin{array}{cccc}
\frac{1}{\sqrt{1-\beta^{2}}} & 0 & \frac{\omega\eta\rho}{\sqrt{1-\beta^{2}}} & 0 \\
0 & 1 & 0 & 0 \\
0 & 0 & \frac{\sqrt{1-\beta^{2}}}{\eta\rho} & 0 \\
0 & 0 & 0 & 1 \\
\end{array}\right),
\label{1.6}
\end{eqnarray}
where we call $\beta=\omega\eta\rho$ in (\ref{1.6}).

The Dirac equation is defined in a curved spacetime background by considering the covariant derivative of a spinor, whose components are defined as $\nabla_{\mu}=\partial_{\mu}+\Gamma_{\mu}\left(x\right)$, where $\partial_{\mu}$ is the partial derivative and $\Gamma_{\mu}\left(x\right)=\frac{i}{2}\,\omega_{\mu ab}\left(x\right)\,\Sigma^{ab}$ is the spinorial connection \cite{naka,bd}. The term $\Sigma^{ab}$ of the spinorial connection is defined as $\Sigma^{ab}=\frac{i}{2}\left[\gamma^{a},\gamma^{b}\right]$. The $\gamma^{a}$ matrices correspond to the standard Dirac matrices in the Minkowski spacetime \cite{greiner}:
\begin{eqnarray}
\gamma^{0}=\hat{\beta}=\left(
\begin{array}{cc}
I & 0 \\
0 & -I \\
\end{array}\right);\,\,\,\,\,\,
\gamma^{i}=\hat{\beta}\,\hat{\alpha}^{i}=\left(
\begin{array}{cc}
 0 & \sigma^{i} \\
-\sigma^{i} & 0 \\
\end{array}\right);\,\,\,\,\,\,\Sigma^{i}=\left(
\begin{array}{cc}
\sigma^{i} & 0 \\
0 & \sigma^{i} \\	
\end{array}\right),
\label{1.7}
\end{eqnarray}
with $\vec{\Sigma}$ and $I$ being the spin vector and the $2\times2$ identity matrix, respectively. The matrices $\sigma^{i}$ are the Pauli matrices and satisfy the relation $\left(\sigma^{i}\,\sigma^{j}+\sigma^{j}\,\sigma^{i}\right)=2\eta^{ij}$. The $\gamma^{\mu}$ matrices are related to the $\gamma^{a}$ matrices via $\gamma^{\mu}=e^{\mu}_{\,\,\,a}\left(x\right)\gamma^{a}$ \cite{bd}. Moreover, the term $\omega_{\mu ab}\left(x\right)$ of the spinorial connection is called the spin connection and can be obtained by solving the Maurer-Cartan structure equations $d\hat{\theta}^{a}+\omega_{\mu\,\,\,b}^{\,\,\,a}\left(x\right)\,dx^{\mu}\wedge\hat{\theta}^{b}=0$ in the absence of torsion \cite{naka}. By solving the Maurer-Cartan structure equations, we obtain the following components of the spinorial connection \cite{bf4,bf16}:
\begin{eqnarray}
\Gamma_{t}&=&-\frac{1}{2}\,\frac{\omega^{2}\eta^{2}\rho}{\sqrt{1-\beta^{2}}}\,\hat{\alpha}^{1}-\frac{i}{2}\,\frac{\omega\eta}{\sqrt{1-\beta^{2}}}\,\Sigma^{3};\nonumber\\
\Gamma_{\rho}&=&\frac{1}{2}\,\frac{\omega\eta}{\left(1-\beta^{2}\right)}\,\hat{\alpha}^{2};\label{1.8}\\
\Gamma_{\varphi}&=&-\frac{1}{2}\frac{\omega\eta^{2}\rho}{\sqrt{1-\beta^{2}}}\,\hat{\alpha}^{1}-\frac{i}{2}\,\frac{\eta}{\sqrt{1-\beta^{2}}}\,\Sigma^{3}.\nonumber
\end{eqnarray}

Hence, by introducing the Dirac oscillator coupling (\ref{1.1}) into the Dirac equation in curved spacetime background, we have \cite{b10}
\begin{eqnarray}
i\gamma^{\mu}\,\partial_{\mu}\psi+i\gamma^{\mu}\,\Gamma_{\mu}\left(x\right)\psi+i\gamma^{\mu}\,m\omega_{0}\rho\,\gamma^{0}\,\delta^{\rho}_{\mu}\,\psi=m\psi.
\label{1.9}
\end{eqnarray}

By using the tetrads given in Eq. (\ref{1.6}) and the components of the spinorial connection given in Eq. (\ref{1.8}), we can write the Dirac equation (\ref{1.9}) in the form \cite{bf16}:
\begin{eqnarray}
m\psi&=&\frac{i}{\sqrt{1-\beta^{2}}}\,\gamma^{0}\,\frac{\partial\psi}{\partial t}+i\gamma^{2}\frac{\omega\eta\rho}{\sqrt{1-\beta^{2}}}\,\frac{\partial\psi}{\partial t}+i\,\gamma^{1}\left(\partial_{\rho}+\frac{1}{2\rho}+m\omega_{0}\,\rho\,\gamma^{0}\right)\psi\nonumber\\
[-2mm]\label{1.10}\\[-2mm]
&+&\frac{i}{\sqrt{1-\beta^{2}}}\frac{\gamma^{2}}{\eta\rho}\frac{\partial\psi}{\partial\varphi}-i\gamma^{2}\frac{\omega^{2}\eta\rho}{\sqrt{1-\beta^{2}}}\,\frac{\partial\psi}{\partial\varphi}+i\,\gamma^{3}\,\frac{\partial\psi}{\partial z}-\frac{\gamma^{0}}{2}\frac{\eta}{\left(1-\beta^{2}\right)}\,\vec{\omega}\cdot\vec{\Sigma}\,\psi.\nonumber
\end{eqnarray}

Observe the presence of the term $\vec{\omega}\cdot\vec{\Sigma}$ in the last term of Eq. (\ref{1.10}) which is related to the rotation-spin coupling as pointed in Ref. \cite{r4}.

From now on, our goal is to to solve the Dirac equation (\ref{1.10}), then, we rewrite it in the form:
\begin{eqnarray}
i\frac{\partial\psi}{\partial t}+i\omega\eta\rho\,\hat{\alpha}^{2}\,\frac{\partial\psi}{\partial t}&=&m\sqrt{1-\beta^{2}}\,\gamma^{0}\,\psi-i\sqrt{1-\beta^{2}}\,\hat{\alpha}^{1}\left(\partial_{\rho}+\frac{1}{2\rho}+m\omega_{0}\,\rho\,\gamma^{0}\right)\psi-i\frac{\hat{\alpha}^{2}}{\eta\rho}\,\frac{\partial\psi}{\partial\varphi}\nonumber\\
[-2mm]\label{1.11}\\[-2mm]
&+&i\omega^{2}\,\eta\rho\,\hat{\alpha}^{2}\,\frac{\partial\psi}{\partial\varphi}-i\sqrt{1-\beta^{2}}\,\hat{\alpha}^{3}\,\frac{\partial\psi}{\partial z}+\frac{1}{2}\frac{\eta\,\vec{\omega}\cdot\vec{\Sigma}}{\left(1-\beta^{2}\right)}\,\psi.\nonumber
\end{eqnarray}

The solution of Eq. (\ref{1.11}) is given in the form: $\psi=e^{-i\mathcal{E}t}\,\left(\phi\,\,\,\chi\right)$, where $\phi$ and $\chi$ are two-spinors. Substituting this solution into (\ref{1.11}), we obtain two coupled equations for $\phi$ and $\chi$, where the first coupled equation is
\begin{eqnarray}
\left[\mathcal{E}-m\sqrt{1-\beta^{2}}-\frac{1}{2}\,\frac{\omega\eta}{\sqrt{1-\beta^{2}}}\,\sigma^{3}\right]\phi&=&-i\sigma^{1}\,\sqrt{1-\beta^{2}}\,\left(\partial_{\rho}+\frac{1}{2\rho}-m\omega_{0}\,\rho\,\right)\chi-i\frac{\sigma^{2}}{\eta\rho}\,\frac{\partial\chi}{\partial\varphi}\nonumber\\
[-2mm]\label{1.12}\\[-2mm]
&+&i\omega^{2}\,\eta\rho\,\sigma^{2}\,\frac{\partial\chi}{\partial\varphi}-i\sqrt{1-\beta^{2}}\,\sigma^{3}\,\frac{\partial\chi}{\partial z}-\omega\eta\mathcal{E}\,\rho\,\sigma^{2}\,\chi,\nonumber
\end{eqnarray}
while the second coupled equation is
\begin{eqnarray}
\left[\mathcal{E}+m\sqrt{1-\beta^{2}}-\frac{1}{2}\,\frac{\omega\eta}{\sqrt{1-\beta^{2}}}\,\sigma^{3}\right]\chi&=&-i\sigma^{1}\,\sqrt{1-\beta^{2}}\,\left(\partial_{\rho}+\frac{1}{2\rho}+m\omega_{0}\,\rho\,\right)\phi-i\frac{\sigma^{2}}{\eta\rho}\,\frac{\partial\phi}{\partial\varphi}\nonumber\\
[-2mm]\label{1.13}\\[-2mm]
&+&i\omega^{2}\,\eta\rho\,\sigma^{2}\,\frac{\partial\phi}{\partial\varphi}-i\sqrt{1-\beta^{2}}\,\sigma^{3}\,\frac{\partial\phi}{\partial z}-\omega\eta\mathcal{E}\,\rho\,\sigma^{2}\,\phi.\nonumber
\end{eqnarray}

In order to solve the two coupled equations (\ref{1.12}) and (\ref{1.13}), we consider the velocity of rotation being small compared with the velocity of the light, that is, $\omega\rho\ll1$. In this way, we can write $\sqrt{1-\beta^{2}}\approx1-\frac{1}{2}\,\beta^{2}+\cdots$. After some calculations, we obtain the following second order differential equation:
\begin{eqnarray}
\left[\left(\mathcal{E}-\frac{\eta\omega}{2}\,\sigma^{3}\right)^{2}-m^{2}\right]\phi&=&-\frac{\partial^{2}\phi}{\partial\rho^{2}}-\frac{1}{\rho}\,\frac{\partial\phi}{\partial\rho}+\frac{1}{4\rho^{2}}\,\phi+\frac{i\sigma^{3}}{\eta\rho}\,\frac{\partial\phi}{\partial\varphi}-m\omega_{0}\,\phi+i\omega^{2}\eta\,\sigma^{3}\,\frac{\partial\phi}{\partial\varphi}\nonumber\\
&-&\mathcal{E}\omega\eta\,\sigma^{3}\,\phi+m^{2}\,\omega_{0}^{2}\,\rho^{2}\,\phi+i\frac{2m\omega_{0}}{\eta}\,\sigma^{3}\,\frac{\partial\phi}{\partial\varphi}+2\omega\eta\,m\omega_{0}\,\rho^{2}\,\sigma^{3}\,\phi\nonumber\\
[-2mm]\label{1.13}\\[-2mm]
&-&\frac{1}{\eta^{2}\rho^{2}}\,\frac{\partial^{2}\phi}{\partial\varphi^{2}}+2\,\omega^{2}\,\frac{\partial^{2}\phi}{\partial\varphi^{2}}+i\,2\omega\mathcal{E}\,\frac{\partial\phi}{\partial\varphi}-\frac{\partial^{2}\phi}{\partial z^{2}}-i\,2m\omega_{0}\rho\,\sigma^{2}\,\frac{\partial\phi}{\partial z}\nonumber\\
&+&\mathcal{O}\left(\beta^{2}\right)+\mathcal{O}\left(m^{-2}\right).\nonumber
\end{eqnarray}

Note in Eq. (\ref{1.13}) that $\phi$ is an eigenfunction of $\sigma^{3}$, whose eigenvalues are $s=\pm1$. Therefore, we label $\sigma^{3}\phi_{s}=\pm\phi_{s}=s\phi_{s}$, where $\phi_{s}=\left(\phi_{+},\,\phi_{-}\right)^{T}$. Furthermore, we can see that the right-hand side of Eq. (\ref{1.13}) commutes with the operators $\hat{J}_{z}=-i\partial_{\varphi}$ \cite{schu} and $\hat{p}_{z}=-i\partial_{z}$, then, we can take the solutions of Eq. (\ref{1.13}) in the form:
\begin{eqnarray}
\phi_{s}=\,e^{i\left(l+\frac{1}{2}\right)\varphi}\,e^{ikz}\,\left(
\begin{array}{c}
R_{+}\left(\rho\right)\\
R_{-}\left(\rho\right)\\	
\end{array}\right),
\label{1.14}
\end{eqnarray}
where $l=0,\pm1,\pm2,...$, $k$ is a constant. Let us consider $k=0$ in order to work with a planar system. Thus, substituting (\ref{1.14}) into (\ref{1.13}), we obtain two noncoupled equations for $R_{+}\left(\rho\right)$ and $R_{-}\left(\rho\right)$ given by
\begin{eqnarray}
\left[\mathcal{E}+\omega\left(l+1/2\right)\right]^{2}R_{s}-m^{2}\,R_{s}&=&-\frac{d^{2}R_{s}}{d\rho^{2}}-\frac{1}{\rho}\,\frac{dR_{s}}{d\rho}+\frac{\zeta_{s}^{2}}{\eta^{2}\rho^{2}}\,R_{s}-2m\omega_{0}\left[s\,\frac{\zeta_{s}}{\eta}+1\right]R_{s}-\omega^{2}\,\zeta_{s}^{2}\,R_{s}\nonumber\\
[-2mm]\label{1.15}\\[-2mm]
&-&2s\,\omega^{2}\,\eta\,\zeta_{s}\,R_{s}-\omega^{2}\eta^{2}\,R_{s}+m^{2}\,\omega_{0}^{2}\,\delta_{s}^{2}\,\rho^{2}\,R_{s}+\mathcal{O}\left(\beta^{2}\right)+\mathcal{O}\left(m^{-2}\right),\nonumber
\end{eqnarray}
where we have defined in (\ref{1.15}) the parameter $\delta_{s}$ and the effective angular momentum $\zeta_{s}$ as
\begin{eqnarray}
\delta_{s}&=&\sqrt{1+2s\,\omega\eta}\nonumber\\
[-2mm]\label{1.16}\\[-2mm]
\zeta_{s}&=&l+\frac{1}{2}\left(1-s\right)+\frac{s}{2}\left(1-\eta\right).\nonumber
\end{eqnarray}

By making a change of variables given by $\xi=m\,\omega_{0}\,\delta_{s}\,\rho^{2}$, we can rewrite the radial equation (\ref{1.15}) in the form:
\begin{eqnarray}
\xi\frac{d^{2}R_{s}}{d\xi}+\frac{dR_{s}}{d\xi}-\frac{\zeta_{s}^{2}}{4\eta^{2}\xi}\,R_{s}-\frac{\xi}{4}\,R_{s}+\frac{\nu_{s}}{4m\omega_{0}\delta_{s}}\,R_{s}=0,
\label{4.14}
\end{eqnarray}
where 
\begin{eqnarray}
\nu_{s}=\left[\mathcal{E}+\omega\left(l+1/2\right)\right]^{2}-m^{2}+2m\omega_{0}\left[s\,\frac{\zeta_{s}}{\eta}+1\right]+\omega^{2}\,\zeta_{s}^{2}+\omega^{2}\eta^{2}+2s\,\omega^{2}\,\eta\,\zeta_{s}.
\label{4.15}
\end{eqnarray} 

A regular solution at the origin is achieved if the solution of the equation (\ref{4.14}) has the form $R_{s}\left(\xi\right)=e^{-\frac{\xi}{2}}\,\xi^{\frac{\left|\zeta_{s}\right|}{2\eta}}\,F_{s}\left(\xi\right)$. Thus, substituting this solution into (\ref{4.14}), we obtain
\begin{eqnarray}
\xi\frac{d^{2}F_{s}}{d\xi^{2}}+\left[\frac{\left|\zeta_{s}\right|}{\eta}+1-\xi\right]\frac{dF_{s}}{d\xi}+\left[\frac{\nu_{s}}{4m\omega_{0}\delta_{s}}-\frac{\left|\zeta_{s}\right|}{2\eta}-\frac{1}{2}\right]F_{s}=0,
\label{4.16}
\end{eqnarray}
which is the confluent hypergeometric equation or the Kummer equation \cite{abra}. The solution of Eq. (\ref{4.16}) regular at the origin is called the Kummer function of first kind, which is given by $F_{s}\left(\xi\right)=\,_{1}F_{1}\left(\frac{\left|\zeta_{s}\right|}{2\eta}+\frac{1}{2}-\frac{\nu_{s}}{4m\omega_{0}\delta_{s}},\frac{\left|\zeta_{s}\right|}{\eta}+1;\xi\right)$. It is well known in the literature that the radial part of the wave function becomes finite everywhere when the parameter $A=\frac{\left|\zeta_{s}\right|}{2\eta}+\frac{1}{2}-\frac{\nu_{s}}{4m\omega_{0}\delta_{s}}$ of the confluent hypergeometric function is equal to a nonpositive integer number, which makes the confluent hypergeometric series to be a polynomial of degree $n$ \cite{landau2}. Thereby, in order to have a wave function being normalized inside the range $0\,<\,\rho\,<\,1/\omega\eta$, we assume that $\sqrt{m\omega_{0}}\ll\omega\eta$. This assumption makes the the amplitude of probability being very small for values where $\rho\,>\,1/\omega\eta$, because we have that $\xi=m\omega_{0}\delta_{s}\rho^{2}\ll1$ when $\rho\rightarrow1/\omega\eta$. Therefore, without loss of generality, we consider the wave function being normalized in the range $0\,<\,\rho\,<\,1/\omega\eta$, since $R_{s}\left(\xi\right)\approx0$ when $\rho\rightarrow1/\omega\eta$. In this way, the relativistic energy levels are
\begin{eqnarray}
\mathcal{E}_{n,\,l}&=&\sqrt{m^{2}+4m\omega_{0}\delta_{s}\left[n+\frac{\left|\zeta_{s}\right|}{2\eta}+\frac{1}{2}\right]-2m\omega_{0}\left[s\,\frac{\zeta_{s}}{\eta}+1\right]-\omega^{2}\,\zeta_{s}^{2}-\omega^{2}\eta^{2}-2s\,\omega^{2}\,\eta\,\zeta_{s}}\nonumber\\
&-&\omega\left[l+1/2\right].
\label{4.17}
\end{eqnarray}
where $n=0,1, 2,...$ and $\zeta_{s}$ and $\delta_{s}$ are given in Eq. (\ref{1.16}). The relativistic energy levels (\ref{4.17}) correspond to the spectrum of energy of the Dirac oscillator under the influence of noninertial effects and the topology of the cosmic string spacetime which are obtained by imposing the condition $\sqrt{m\omega_{0}}\ll\omega\eta$. 

Moreover, we have that the effects of curvature break the degeneracy of the relativistic energy levels of the Dirac oscillator (\ref{4.17}). By taking the limit $\eta\rightarrow0$, we have the spectrum of energy of the Dirac oscillator in the Minkowski spacetime under the influence of noninertial effects of the rotating frame (\ref{1.6}). Recently, we have studied the influence of the noninertial effects of the Fermi-Walker reference frame on the Dirac oscillator in the cosmic string spacetime \cite{b10}. By comparing the results obtained in Ref. \cite{b10} with the relativistic energy levels (\ref{4.17}), we have that new contributions to the energy levels arise from the noninertial effects of the rotating frame (\ref{1.6}). This stems from the fact that the local reference frame of the observers (\ref{1.6}) is not defined in the rest frame of the observers at each instant $\left(\hat{\theta}^{0}\neq e^{0}_{\,\,\,t}\left(x\right)\,dt\right)$ in contrast to the Fermi-Walker reference frame \cite{misner,r5}. Finally, we have that in last term of Eq. (\ref{4.17}) the coupling between the quantum number $l$ and the angular velocity $\omega$.

Let us discuss the nonrelativistic limit of the energy levels (\ref{4.17}). The nonrelativistic limit of Eq. (\ref{4.17}) can be obtained by applying the Taylor expansion up to the first order terms. In this way, the nonrelativistic limit of the energy levels (\ref{4.17}) are
\begin{eqnarray}
\mathcal{E}_{n\,l}\approx m+2\omega_{0}\delta_{s}\left[n+\frac{\left|\zeta_{s}\right|}{2\eta}+\frac{1}{2}\right]-\omega_{0}\left[s\,\frac{\zeta_{s}}{\eta}+1\right]-\frac{\omega^{2}\,\zeta_{s}^{2}}{2m}-\frac{\omega^{2}\eta^{2}}{2m}-s\,\frac{\omega^{2}\,\eta\,\zeta_{s}}{m}-\omega\left[l+\frac{1}{2}\right].
\label{4.18}
\end{eqnarray}

The first term of the nonrelativistic energy levels (\ref{4.18}) corresponds to the rest energy of the quantum particle. The remaining terms of the energy levels (\ref{4.18}) corresponds to the energy levels of a harmonic oscillator under the influence of noninertial effects of the rotating frame (\ref{1.6}) and the topology of a disclination \cite{furt2}. By taking $\omega\rightarrow0$, we recover the results of Ref. \cite{furt2}, but for a spin-$1/2$ quantum particle. Hence, we also have that influence of the topological defect breaks the degeneracy of the energy levels of the harmonic oscillator as pointed out in Ref. \cite{furt2}. In the limit $\eta\rightarrow0$, we have the spectrum of energy of the harmonic oscillator in the absence of defects, but under the influence of noninertial effects of the rotating frame (\ref{1.6}). Besides, we have the presence of the coupling between the quantum number $l$ and the angular velocity $\omega$, which corresponds to the Page-Werner {\it et al} term \cite{r1,r2}.

Hence, the bound states solutions of the Dirac oscillator in the rotating frame (\ref{1.6}) in the cosmic string spacetime is obtained by assuming the condition $\sqrt{m\omega_{0}}\ll\omega\eta$. Without this condition, we cannot consider the amplitude of probability of finding the Dirac neutral particle in the non-physical region of the spacetime being very small.

On the other hand, let us discuss the behaviour of the Dirac oscillator without assuming that $\sqrt{m\omega_{0}}\ll\omega\eta$. Returning to the solution of Eq. (\ref{4.16}), we have that by imposing the condition where he confluent hypergeometric series becomes a polynomial of degree $n$, then, the radial wave function becomes finite everywhere including the non-physical region of the spacetime established in Eq. (\ref{1.5}). Thereby, a normalized wave function can be obtained if we impose first $m\omega_{0}$ is very small. The assumption  $m\omega_{0}\ll1$ allows us to consider a fixed radius $\rho_{0}=1/\omega\eta$ and a fixed value for the parameter of the confluent hypergeometric function $B=\frac{\left|\zeta_{s}\right|}{\eta}+1$ in such a way that the parameter of the confluent hypergeometric function $A=\frac{\left|\zeta_{s}\right|}{2\eta}+\frac{1}{2}-\frac{\nu_{s}}{4m\omega_{0}\delta_{s}}$ can be considered large, without loss of generality. In this way, we can write the confluent hypergeometric function in the form \cite{abra,b10,b12}:
\begin{eqnarray}
_{1}F_{1}\left(A,\,B,\,\xi_{0}=m\,\omega_{0}\,\delta_{s}\,\rho^{2}_{0}\right)&\approx&\frac{\Gamma\left(B\right)}{\sqrt{\pi}}\,e^{\frac{\xi_{0}}{2}}\left(\frac{B\xi_{0}}{2}-A\xi_{0}\right)^{\frac{1-B}{2}}\times\nonumber\\
&\times &\cos\left(\sqrt{2B\xi_{0}-4A\xi_{0}}-\frac{B\pi}{2}+\frac{\pi}{4}\right),
\label{4.19}
\end{eqnarray}
where $\Gamma\left(B\right)$ is the gamma function. Hence, our last step in order to obtain a normalized wave function in the range $0\,<\,\rho\,<\,1/\omega\eta$ is to impose that the radial wave function vanishes at $\rho\rightarrow1/\omega\eta$, that is,
\begin{eqnarray}
R_{s}\left(\xi_{0}\right)=R_{s}\left(m\omega\delta_{s}\rho_{0}^{2}\right)=0,
\label{4.20}
\end{eqnarray}
where $\rho_{0}=1/\omega\eta$. Therefore, by writing the radial wave function $R_{s}\left(\xi\right)=e^{-\frac{\xi}{2}}\,\xi^{\frac{\left|\zeta_{s}\right|}{2\eta}}\,_{1}F_{1}\left(A,\,B,\,\xi\right)$ in terms of the expression given in Eq. (\ref{4.19}) and by applying the boundary condition (\ref{4.20}), we obtain
\begin{eqnarray}
\mathcal{E}_{n,\,l}&\approx&\sqrt{m^{2}+\frac{1}{\rho_{0}^{2}}\left[n\pi+\frac{\left|\zeta_{s}\right|}{2\eta}\,\pi+\frac{3\pi}{4}\right]^{2}-2\,m\omega_{0}\left[s\,\frac{\zeta_{s}}{\eta}+1\right]-\omega^{2}\,\zeta_{s}^{2}-\omega^{2}\eta^{2}-2s\,\omega^{2}\,\eta\,\zeta_{s}}\nonumber\\
&-&\omega\left[l+\frac{1}{2}\right],
\label{4.21}
\end{eqnarray} 
where $n=0,1, 2,...$ and the parameters $\zeta_{s}$ and $\delta_{s}$ are defined in Eq. (\ref{1.16}).

The relativistic energy levels (\ref{4.21}) correspond the spectrum of energy of the Dirac oscillator under the influence of noninertial effects of the rotating frame (\ref{1.6}) and the topology of the cosmic string spacetime by imposing the the conditions of vanishing the radial wave function at $\rho\rightarrow1/\omega\eta$ and $m\omega_{0}\ll1$. This corresponds to the spectrum of energy of the Dirac oscillator under the influence of a hard-wall confining potential induced by noninertial effects. In this case, the geometry of the spacetime plays the role of a hard-wall confining potential due to the presence of noninertial effects that restricts the physical region of the spacetime \cite{b10,b12}. Observe the difference between the relativistic energy levels obtained in Eqs. (\ref{4.17}) and (\ref{4.21}) stems from the restriction of the physical region of the spacetime imposed by noninertial effects and the conditions imposed on the Dirac oscillator frequency $\omega_{0}$. We can also see that the curvature effects on the relativistic energy levels (\ref{4.21}) change the degeneracy of the spectrum of energy. Besides, by taking the limit $\eta\rightarrow1$, the effects of curvature vanish and we obtain that Eq. (\ref{4.21}) is the relativistic spectrum of energy of the Dirac oscillator under the influence of noninertial effects of the rotating frame (\ref{1.6}) in the Minkowski spacetime. Again, we can see that the noninertial effects of the rotating frame (\ref{1.6}) yield new contributions to the relativistic energy levels (\ref{4.21}) in contrast to the results of Ref. \cite{b10} obtained in the Fermi-Walker reference frame \cite{misner}.

Finally, let us obtain the nonrelativistic limit of the energy levels (\ref{4.21}). By applying the Taylor expansion, then, the nonrelativistic limit of the energy levels (\ref{4.21}) are
\begin{eqnarray}
\mathcal{E}_{n\,l}&\approx& m+\frac{1}{2m\rho_{0}^{2}}\left[n\pi+\frac{\left|\zeta_{s}\right|}{2\eta}\,\pi+\frac{3\pi}{4}\right]^{2}-\omega_{0}\left[s\frac{\zeta_{s}}{\eta}+1\right]-\frac{\omega^{2}\,\zeta_{s}^{2}}{2m}-\frac{\omega^{2}\eta^{2}}{2m}\nonumber\\
&-&s\,\frac{\omega^{2}\,\eta\,\zeta_{s}}{m}-\omega\left[l+\frac{1}{2}\right],
\label{4.22}
\end{eqnarray}
where the first term of the nonrelativistic energy levels (\ref{4.22}) corresponds to the rest energy of the quantum particle. 

The nonrelativistic energy levels (\ref{4.22}) correspond to the spectrum of energy of a harmonic oscillator under the influence of a hard-wall confining potential induced by noninertial effects and the topology of a disclination \cite{kat,furt,furt2}. We can also observe that the degeneracy of the energy levels of the harmonic oscillator is broken by the topology of the defect. By taking the limit $\eta\rightarrow1$, we obtain the spectrum of energy of the harmonic oscillator under the influence of a hard-wall confining potential induced by noninertial effects in the absence of defects. Furthermore,  we have that the energy levels (\ref{4.22}) are proportional to $n^{2}$ in contrast to the previous result given in Eq. (\ref{4.18}), whose spectrum of energy is proportional to the quantum number $n$. However, in both cases, we have the presence of the Page-Werner {\it et al} term \cite{r1,r2}, which is the coupling between the quantum number $l$ and the angular velocity $\omega$ given by the last term of Eq. (\ref{4.22}).

\section{conclusions}
 
In this paper, we have studied the influence of noninertial effects of a rotating frame and the topology of the cosmic string spacetime on the Dirac oscillator. We have shown that both the topology of the cosmic string spacetime and the noninertial effects restrict the physical region of the spacetime where the quantum particle can be placed. Moreover, by analysing the behaviour of the oscillatory frequency, we have seen that two different bound states solutions for the Dirac equation can be obtained.  
 
We have shown by assuming $\sqrt{m\omega_{0}}\ll\omega\eta$ that the amplitude of the wave function becomes very small for values of the radial coordinate given by $\rho>1/\omega\eta$, therefore, the wave function could be considered to be normalized inside the physical region of the spacetime defined in the range $0\,<\,\rho\,<\,1/\omega\eta$. As a consequence of the assumption $\sqrt{m\omega_{0}}\ll\omega\eta$, we have obtained the the relativistic spectrum of energy of the Dirac oscillator under the influence of the noninertial effects of a rotating frame and the topology of the cosmic string spacetime. Moreover, we have seen that the effects of curvature concentrated on the cosmic string axis break the degeneracy of the spectrum of energy of the Dirac oscillator.
 
On the other hand, without assuming the condition $\sqrt{m\omega_{0}}\ll\omega\eta$, we have seen that we cannot normalize the radial wave function by imposing that the confluent hypergeometric series becomes a polynomial of degree $n$, because the radial wave function becomes finite either in the physical region of the spacetime and in the non-physical region of the spacetime. Thereby, we have shown that a normalized wave function can be obtained by assuming that $m\omega_{0}$ is very small and by imposing that the radial wave function vanishes at $\rho\rightarrow1/\omega\eta$. In this case, we have obtained the relativistic energy levels that corresponds to the spectrum of energy of the Dirac oscillator under the influence of a hard-wall confining potential induced by noninertial effects, where the geometry of the spacetime plays the role of a hard-wall confining potential \cite{b10,b12}. Again, we have observed that the effects of curvature concentrated on the cosmic string axis change the degeneracy of the spectrum of energy of the Dirac oscillator under the influence of a hard-wall confining potential induced by noninertial effects.

Finally, we have discussed in all cases above the nonrelativistic limit of the energy levels. By assuming the condition $\sqrt{m\omega_{0}}\ll\omega\eta$, we have seen that the nonrelativistic energy levels correspond to the harmonic oscillator spectrum under the influence of noninertial effects of the rotating frame (\ref{1.6}) and the topology of a disclination \cite{furt2}. Without assuming the condition $\sqrt{m\omega_{0}}\ll\omega\eta$, but by considering $m\omega_{0}\ll1$ and by imposing that the radial wave function vanishes at $\rho\rightarrow1/\omega\eta$, we have seen that the nonrelativistic energy levels (\ref{4.22}) correspond to the harmonic oscillator spectrum under the influence of a hard-wall confining potential induced by noninertial effects and the topology of a disclination \cite{kat,furt,furt2}.

\acknowledgments

The author would like to thank CNPq (Conselho Nacional de Desenvolvimento Cient\'ifico e Tecnol\'ogico - Brazil) for financial support.

\end{document}